\documentclass[doublecol,aps,epl,preprintnumbers,nofootinbib,floatfix]{epl2}

\usepackage{graphicx,color,amsmath}

\newcommand{\D}{\mathrm{d}}
\newcommand{\e}{\mathrm{e}}
\newcommand{\half}{\frac{1}{2}}
\newcommand{\be}{\begin{equation}}
\newcommand{\ee}{\end{equation}}
\newcommand{\bea}{\begin{eqnarray}}
\newcommand{\eea}{\end{eqnarray}}

\newcommand{\kbt}{k_{\mathrm{B}}T}

\newcommand{\lb}{l_\mathrm{B}}

\newcommand{\kd}{\kappa_\mathrm{D}}

\begin{document}


\title{Surface Tension of Electrolyte Solutions: \\
A Self-consistent Theory}
\shorttitle{Surface Tension of Electrolyte Solutions: A Self-consistent Theory}
\author{Tomer Markovich \inst{1} \and David Andelman \inst{1} \and Rudi Podgornik \inst{2}}
\shortauthor{T. Markovich \etal}

\institute{ \inst{1} Raymond and Beverly Sackler School of Physics and Astronomy Tel Aviv
University, Ramat Aviv, Tel Aviv 69978, Israel. \\
\inst{2} Department of Theoretical Physics, J. Stefan Institute, and Department of Physics, Faculty of Mathematics and Physics, University of Ljubljana, 1000 Ljubljana, Slovenia.}
\date{April 3, 2014}

\abstract{
We study the surface tension of electrolyte solutions at the air/water and oil/water interfaces. Employing field-theoretical methods and considering short-range interactions of anions with the surface, we expand the Helmholtz free energy to first-order in a loop expansion and calculate the excess surface tension. Our approach is self-consistent and yields an analytical prediction that reunites the Onsager-Samaras pioneering result (which does not agree with experimental data), with the ionic specificity of the Hofmeister series. We obtain analytically the surface-tension dependence on the ionic strength, ionic size and ion-surface interaction, and show consequently that the Onsager-Samaras result is consistent with the one-loop correction beyond the mean-field result. Our theory fits well a wide range of concentrations for different salts using one fit parameter, reproducing the reverse Hofmeister series for anions at the air/water and oil/water interfaces.}

\maketitle

\section{Introduction}

When salts are added in small quantities to an aqueous solution, its surface tension generally increases~\cite{Adamson,Pugh}.
Wagner~\cite{Wagner} was the first to connect this finding with the dielectric discontinuity at the air/water surface,
suggesting dielectric image interactions  as a possible explanation.
This idea was implemented in the pioneering work of Onsager and Samaras (OS) that
was built upon the work of Debye and H\"uckel~\cite{Debye1923}. In their model, OS found a universal limiting law
for the dependence of the excess surface tension on the salt concentration~\cite{onsager_samaras}.
However, the OS result implies an {increase in the surface tension that is independent of the ion type},
which  turned out to be violated in many physical realizations~\cite{Kunz_Book}.
This led to numerous investigations of non-electrostatic {\it ion-specific} interactions between
ions and surfaces~\cite{Dan2011,Kunz_Book}, and their role in modifying
surface tension of electrolyte solutions~\cite{Pugh}.
However, even nowadays a fundamental understanding of surface tension of electrolyte solutions is still missing.

On a broader scope, ion-specific effects date back to the late 19th century, when Hofmeister~\cite{hofmeister}
measured the amount of protein precipitation from solution in presence of various salts, and found a
universal (Hofmeister) series of ionic activity. The same ionic series emerged in a large
variety of chemical and biological experiments~\cite{collins1985,ruckenstein2003a,kunz2010},
such as forces between mica or silica surfaces~\cite{Sivan2009,Sivan2013,pashely},
osmotic pressure in the presence of (bio)macromolecules~\cite{parsegian1992,parsegian1994}, and quite notably in the surface tension measurements at the air/water and oil/water interfaces~\cite{air_water_2,air_water_3}.
For simple monovalent salts the air/water surface tension depends strongly on the type of anion,
while the dependence on the cation type is weaker~\cite{Frumkin1924}, and
is consistent with the fact that anion concentration exceeds that of cations at the air/water interface.
For halide ions, the lighter ones lead to a larger excess in surface tension in
a sequence that is precisely the {\it reverse} of the Hofmeister series.

The OS treatment of electrolyte surface tension attracted much interest and generated
a vast number of modifications to the original model, in particular more recent ones~\cite{dean2004,dean2003,levin200x,levin200y,Netz2010,Netz2012,Netz2013} that are relevant for
our approach advocated below.
Specifically, Dean and Horgan~\cite{dean2004} calculated the
ionic solution surface-tension to first order in a cumulant
expansion, where the zeroth-order is equivalent to
the Debye-H\"uckel approximation~\cite{Debye1923}.  The surface-specific
interactions were included via an ionic surface-exclusion
layer in the vicinity of the dielectric interface. In another study,
a surface cation-specific, short-range interaction
was added~\cite{dean2003}, but the corresponding surface tension was
not calculated.

Levin and coworkers~\cite{levin200x,levin200y} calculated the solvation free-energy
of polarizable ions at air/water and oil/water interfaces.
Their model relies on three surface-ion terms that are added in an {\it ad hoc} way to the PB equation:
image charge interaction in the presence of a Stern layer, ionic cavitation energy and ionic polarizability.
Although these three terms give a clue to the origin of the
ion-surface interaction, they cannot be  added in a self-consistent way.
Furthermore, these terms are neither completely independent nor
can be obtained from a mean-field (MF) theory~\cite{Kunz_Book,hofmeister}.
Double-counting of electrostatic contributions is a common ambiguity as manifested by the  image charge term.  This term cannot simply be added to the mean potential in the Boltzmann weight factor, as it is not a solution of the Poisson-Boltzmann (PB) equation
close to a planar wall. The PB equation is a MF equation that follows from minimization of a certain free-energy functional.
A consistent way to generalize it would have to be  based on an augmented free-energy functional that would then give the generalized PB equation. As will be shown below, within a {\it self-consistent} treatment, the image charge term follows from the one-loop correction to the MF result~\cite{podgornik1988}.

A different line of reasoning was initiated by Netz and coworkers~\cite{Netz2010,Netz2012,Netz2013},
who calculated the surface tension for both charged and neutral surfaces by
combining molecular dynamics (MD) with the PB theory on a MF level.
Their results fit well the experiments performed with hydrophobic and hydrophilic surfaces and agree
with the Hofmeister series results.

An inclusion of ion-specific effects in a theory of electrolyte
solutions is highly desirable but also quite difficult to obtain rigorously.
We believe that a very promising line of reasoning lies within a phenomenological approach~\cite{Dan2011},
where short-range non-electrostatic interactions are explicitly added to the electrostatic free energy.
This approach allows a clearcut separation between various degrees of freedom.
The additional free-energy terms describe specific electrolyte features that go beyond regular PB theory
and depend on the ion chemical nature, size, charge, polarizability, and the preferential
ion-solvent interaction~\cite{Dan2011,US1,US2,Onuki_curr_op,delacruz2012}.

In this Letter, we propose such a phenomenological approach that not only describes successfully the surface tension of electrolyte solutions,
but also adds instructive insight into the corresponding surface interaction parameters.
Our approach is self-consistent and yields an analytical prediction that reunites the Onsager-Samaras pioneering result, which does not agree with experimental data, with the ionic specificity of the Hofmeister series.
We take ionic specificity into account through the ionic size and a
short-range ion-surface interaction~\cite{levin200x,levin200y},
characterized by a single phenomenological adhesivity parameter~\cite{diamant1996,dean2003}.
Our theory is formally consistent, fits well a variety of experimental interfacial tension data at the
air/water and oil/water interfaces and reproduces the revered Hofmeister series
for several types of monovalent anions.

\section{Model}

In our model the water and air phases are taken as two continuum media with uniform dielectric
constant $\varepsilon_{\rm w}$ and $\varepsilon_{\rm a}$, respectively, and with a sharp planar boundary between them at $z=0$.
The water volume $V=AL$ is modelled as a box of cross-section $A$ and an arbitrary large length, $L\to \infty$.
We consider a monovalent symmetric ($1$:$1$) salt, where the ions carry a unit charge $\pm e$ and are taken to be point-like.

We also include explicitly an ion-surface interaction that is short ranged with a scale of the order of the ionic size, denoted as $a$. As it is experimentally known that anions are less hydrated than their cation counter-parts~\cite{Frumkin1924},
the ion-surface interaction is considered only for anions, but
the model can easily be implemented for more general setups to also include the cation-surface interaction~\cite{tobepub}.

The model Hamiltonian is:
\begin{equation}
\label{e1}
H = \half\sum_{i\ne j}q_i q_j u(r_i,r_j)
+ \alpha a\sum_{i\, \epsilon\, {\rm anions}}\delta\left(z_i\right) ,
\end{equation}
where $q_i=\pm e$ are the cation/anion electric charge, and the adhesivity parameter $\alpha$ is expressed in units of $\kbt$.
The first term is the usual Coulombic interaction $\nabla^2 u(r)=-\frac{4\pi}{\varepsilon_{\rm w}}\delta(r)$, where the diverging self-energy of the ions should be subtracted. The second term is a short-range ion-surface interaction modeled as a Dirac $\delta$-function, akin to the adhesivity used by Davies~\cite{Davies}. Its strength is parameterized by the parameter $\alpha$. As ions have a finite size, we defined $a$ as their minimal distance of approach, which will serve hereafter as the shortest length cutoff. The limit $a \to 0$ is appropriate for point-like ions. Note that we have chosen to use the same finite size $a$ for the ionic minimal distance of approach
as well as for the range of the ion-surface interaction.

Using the standard Hubbard-Stratonovich transformation ~\cite{podgornik1988}, the grand-partition function
(up to a normalization factor) can be written as
\begin{eqnarray}
\label{e2}
\Xi = \left({\det[\beta^{-1} u(r,r')]}\right)^{-1/2}\int D\phi\,\e^{-S\left[\phi\right]}\, ,
\end{eqnarray}
with $\beta=1/\kbt$ being the inverse thermal energy and $S$ plays the role of a ``field action",
\begin{eqnarray}
\label{e3}
\nonumber& S &= \int_{V}\D^3r\,\left(\frac{\beta\varepsilon_{\rm w}}{8\pi}[\nabla \phi(r)]^2
 - \lambda_{+}\e^{-i\beta e\phi(r)} - \lambda_{-}\e^{i\beta e\phi(r)}\right) \\
& &- a\lambda_{-}\left( \e^{- \beta\alpha} - 1\right)\int_{A}\D^2r\,\e^{i\beta e\phi(z{=}0)}  \, .
\end{eqnarray}
The surface term accounts only for the anion (charge $-e$) interaction and the fugacities $\lambda_\pm$
are defined via the chemical potentials $\mu_\pm$ as:
$\lambda_{\pm} \equiv a^{-3}\exp\left[\beta\mu_{\pm} + \half\beta e^2u(r,r)\right]$.
In the surface term of eq.~(\ref{e3}) we also make use of the discretized $\delta$-function property, $a\delta(0)=1$, giving
\begin{equation}
\label{e4}
\e^{-\beta\alpha a\delta\left(z\right)}=1+a\left(\e^{-\beta\alpha}-1\right)\delta(z) \, .
\end{equation}

By rescaling the action into a dimensionless form, $S\to S/{\cal C}$,  we introduce the coupling constant~\cite{tobepub},
${\cal C}=2\pi\lb^2 a\lambda_{-}\left({\rm e}^{-\beta\alpha}-1\right)$, where
$\lb = \beta e^2/\varepsilon_{\rm w}$ is the Bjerrum length.
In the weak-coupling regime, ${\cal C}\ll 1$, the field action of eq.~(\ref{e3}) can be expanded around its MF solution $S_0$ as $S\simeq S_0+S_1+...$, where $S_1$ is the one-loop correction. The grand-potential can then be written to the one-loop order
as~\cite{podgornik1989}:
\begin{eqnarray}
\label{e5}
\Omega = \Omega_0 + \Omega_1 = \kbt \left[ S_0
+ \half {\rm Tr}\ln H_2(r,r') \right]  ,
\end{eqnarray}
{with $H_2(r,r') = \left[\frac{\delta^2S}{\delta\phi(r)\delta\phi(r')}\right]_0$ being the Hessian of $S$ evaluated at its MF value.
The MF solution, corresponding to the saddle-point of $S$, can be rewritten in the standard form of a PB equation for the mean-field electrostatic potential $\psi$ by identifying $\psi = i\phi_{\rm MF}$}.

The electrostatic potential in the air $(z<0)$, $\psi_1(z)$, satisfies the Laplace equation, while the electrostatic potential in the aqueous phase $(z>0)$, $\psi_2(z)$, obeys the PB equation in the form
\begin{eqnarray}
\label{e6}
& &\psi_2^{\prime\prime}(z) = \frac{8\pi e n_b}{\varepsilon_{\rm w}} \sinh\left(\beta e\psi_2\right) \, .
\end{eqnarray}
We used the electro-neutrality condition, $\sum_{i=\pm} \lambda_iq_i=0$, which for symmetric electrolytes implies,
$\lambda_{\pm}=\lambda$, and on the MF level $\lambda=n_b$ where $n_b$ is the salt bulk concentration.
The boundary condition is:
\begin{equation}
\varepsilon_{\rm w}\psi_2^\prime|_{_{0^+}} - \varepsilon_{\rm a} \psi_1^\prime|_{_{0^-}} = -4\pi \sigma_{0}\e^{\beta e\psi_0}\,,
\end{equation}
where the RHS of the above equation represents an effective surface charge, $\sigma_{0}\e^{\beta e\psi_0}$,
induced by the surface potential $\psi_0\equiv\psi(z{=}0)$, and $\sigma_{0}$ is
\begin{equation}
\label{e7}
\sigma_{0} = -an_b e\left(\e^{-\beta\alpha}-1\right)\, .
\end{equation}
Note that for $\alpha>0$ (repulsive interaction), the surface charge $\sigma_0$ is positive.
In the weak-coupling MF regime, ${\cal C}\ll 1$ implies that $\sigma_{0}$ is small and the surface potential $\psi_0$ is also found self-consistently to be weak.
Therefore, $\beta e\psi(z)\ll 1$ and the above PB equation (\ref{e6}) can be linearized.
Using the fact that the electrostatic field vanishes at $z\rightarrow\pm\infty$, $\psi_{1,2}$ become:
\begin{eqnarray}
\label{e8}
\nonumber && \psi_1 = \psi_0 = \frac{4\pi\sigma_{0}}{\kd\varepsilon_{\rm w} - 4\pi\beta e\sigma_{0}} \\
&& \psi_2 = \psi_0\e^{-\kd z}\, ,
\end{eqnarray}
where $\kd^{-1} \equiv \left(8\pi n_b\lb\right)^{-1/2}$ is the Debye length.

$\Omega_0$  is  obtained by substituting eq.~(\ref{e8}) into eq.~(\ref{e5}),
while the fluctuation contribution  around the MF, $\Omega_1$, can be obtained
by using the argument principle~\cite{podgornik1989,attard}.
The discrete sum of the Hessian operator eigenvalues is expressed in terms
of its secular determinant, $D_{\nu}$, yielding
\be
\label{e9}
\Omega_1=\frac{A\kbt}{8\pi^2} \int\D^2k\,\ln\left(\frac{D_{\nu=1}(k)}{D_{\nu=0}(k)}\right) \, ,
\ee
and the integral is over the transverse wavevector ${\bf k} = (k_x,k_y)$.
The index $\nu$ refers to the eigenvalue equation of the form
\begin{equation}
\label{e10}
f_{\nu}''(z) - k^2f_{\nu}(z)=\nu \kd^2\cosh(\beta e\psi_0\e^{-\kd z})f_{\nu}(z),
\end{equation}
with the boundary condition at $z\to 0$
\begin{eqnarray}
\label{e11}
& & \varepsilon_{\rm w} f_{\nu}'(0^+) - \varepsilon_{\rm a} f_{\nu}'(0^-) = \omega_{\nu} f_{\nu}(0)\, ,
\end{eqnarray}
and $ f_{\nu}(z{\rightarrow}\pm\infty) = 0$, where  $\omega_{\nu}=-4\pi\nu\beta e\sigma_{0}\e^{\beta e\psi_0}$.
The secular determinant, $D_{\nu}$, is obtained from the boundary conditions.

In the weak-coupling regime, the RHS of eq. (\ref{e10}) can be linearized, yielding
\begin{eqnarray}
\label{e11b}
f_{\nu}''(z) - p_\nu^2f_{\nu}(z) = 0\, ,
\end{eqnarray}
where $p_{\nu}^2 \equiv k^2 + \nu\kd^2$. The general solution of the above equation is:
\begin{eqnarray}
\label{s1}
\nonumber & &f_1(z) = A_1\e^{kz}\\
& &f_2(z) = A_2C_{\nu}(z) + B_2S_{\nu}(z) \, ,
\end{eqnarray}
where the subscript `$1$' stands for the $z<0$ region (air), and `$2$' stands for the $z>0$ region (water).
The functions $C_{\nu}(z)$ and $S_{\nu}(z)$ are defined as the even and odd solutions for $z>0$, respectively:
\begin{eqnarray}
\label{s2}
\nonumber& &C_{\nu}(0) = 1 \quad ; \quad C'_{\nu}(0) = 0 \\
& &S_{\nu}(0) = 0 \quad ; \quad S'_{\nu}(0) = 1 \, ,
\end{eqnarray}
and equal to $C_\nu \equiv \cosh(p_{\nu}z)$ and $S_\nu \equiv \sinh(p_{\nu}z)/p_\nu$. The
derivatives with respect to $z$ are denoted by $C'$ and $S'$.

We use the method described in Refs.~\cite{podgornik1989,attard,fdet} in order to find the secular determinant, $D_\nu$:
\begin{eqnarray}
\label{s3}
D_{\nu} = \det\left[M
                    + N \begin{pmatrix}
                            C_{\nu}(L) & S_{\nu}(L) \\
                            C'_{\nu}(L) & S'_{\nu}(L)
                        \end{pmatrix}
                        \right] \, .
\end{eqnarray}
The boundary conditions of the eigenvalue equation, eq.~(11), can be written
using the matrices $M$ and $N$ that satisfy:
\begin{eqnarray}
\label{s4}
M   \begin{pmatrix}
        u_{\nu}(0) \\
        u'_{\nu}(0)
    \end{pmatrix} + N \begin{pmatrix}
                        u_{\nu}(L) \\
                        u'_{\nu}(L)
                      \end{pmatrix} = 0 \, ,
\end{eqnarray}
while $u_{\nu}$ is equal to either $C_{\nu}$ or $S_{\nu}$.
We can choose conveniently these matrices to be:
\begin{eqnarray}
\label{s5}
M = \begin{pmatrix}
        -\omega_{\nu} - \varepsilon_{\rm a} k &  \varepsilon_{\rm w} \\
        0 & 0
    \end{pmatrix} \quad ; \quad
N = \begin{pmatrix}
        0 & 0 \\
        0 & 1
    \end{pmatrix} \, ,
\end{eqnarray}
and the resulted secular determinant is:
\begin{eqnarray}
\label{s6}
\nonumber D_{\nu} \!\!\!\! &=& \!\!\!\!{\rm det}
\begin{pmatrix}
    -\omega_{\nu} - \varepsilon_{\rm a} k &  \varepsilon_{\rm w} \\
    C'_{\nu}(L) & S'_{\nu}(L)
\end{pmatrix} \\
&=& \!\!\!\!-\left( \omega_{\nu} + \varepsilon_{\rm a} k \right)\cosh(p_{\nu}L) - p_{\nu}\varepsilon_{\rm w} \sinh(p_{\nu}L) \, . \end{eqnarray}
The leading asymptotic term in the limit, $L\rightarrow\infty$, leads to:
\begin{eqnarray}
\label{s7}
D_{\nu} = -\half\left[\omega_{\nu} + \varepsilon_{\rm w} p_{\nu} + \varepsilon_{\rm a} k \right]\e^{p_{\nu}L} \, ,
\end{eqnarray}

Inserting $D_{\nu}$ into eq.~(\ref{e9}) and expressing $\kd$ in terms of the fugacity gives:
\begin{eqnarray}
\label{e12}
\nonumber & \Omega_1 &= \frac{V\kbt}{12\pi }\left[\left(\Lambda^2+\kd^2\right)^{3/2} - \kd^3 - \Lambda^3 \right] \\
& &+ \frac{A\kbt}{4\pi }\int_0^{\Lambda}\D k k \ln\left(\frac{\varepsilon_{\rm w} p + \varepsilon_{\rm a} k + \omega}{\left(\varepsilon_{\rm w} + \varepsilon_{\rm a} \right)k}\right) \, ,
\end{eqnarray}
where $\omega \equiv \omega_{\nu = 1}$ and $p\equiv p_{\nu=1}$.
Although Coulombic interactions between point-like ions diverge at zero distance,
in reality such a divergence is avoided because of steric repulsion for finite size ions.
A common way in field theory to avoid this issue without introducing an explicit
steric repulsion is to employ a short length (UV) cutoff. For isotropic two-dimensional integrals,
as in eq.~(\ref{e12}) above, the UV cutoff is taken to be $\Lambda = 2\sqrt{\pi}/a$.
As noted after eq.~(1), $a$ is the average minimal distance between ions, and thus is related indirectly to the ion size.

\begin{figure}[h!]
\center
\includegraphics[width=0.45\textwidth]{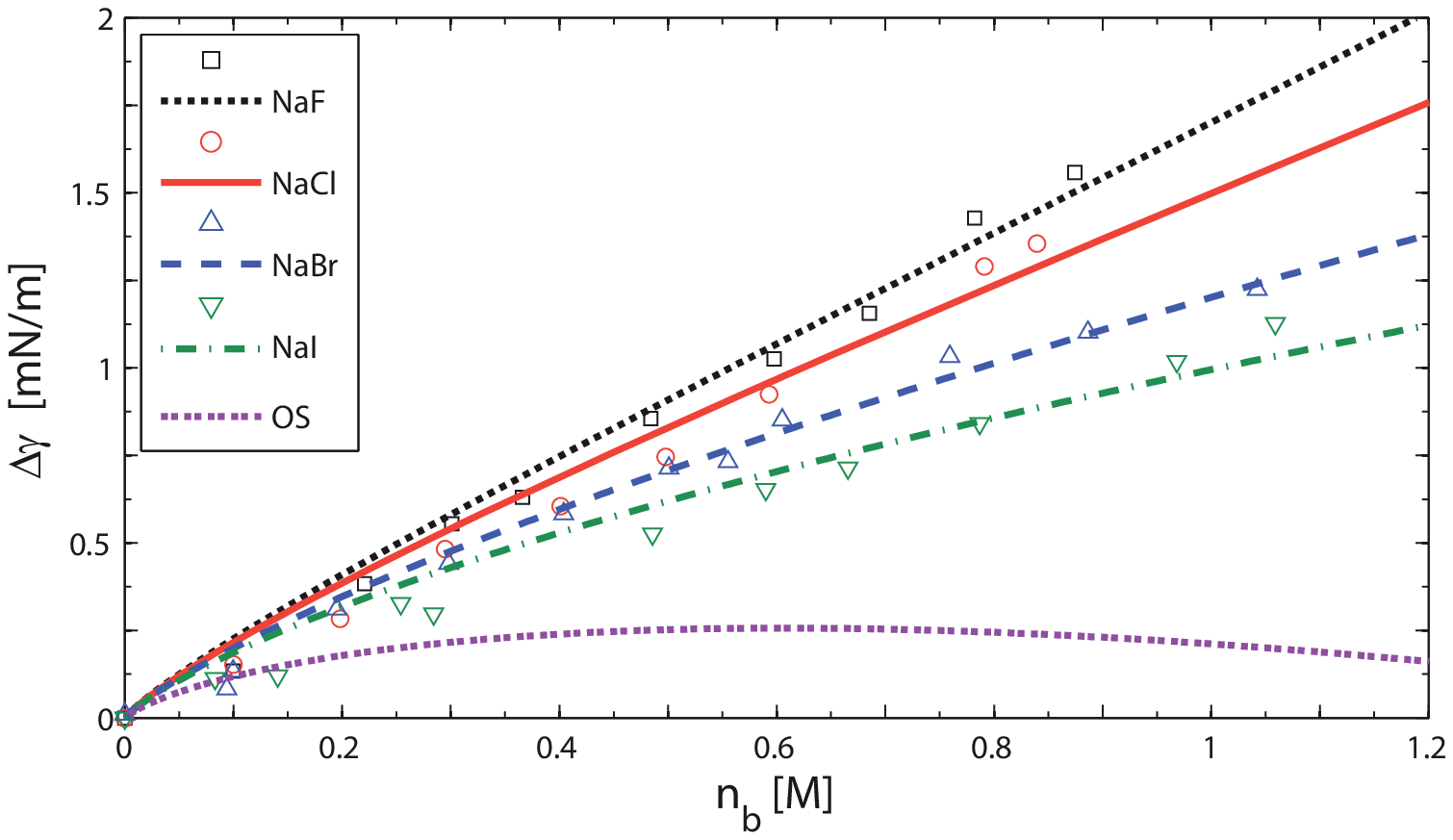}  
\caption{\textsf{(color online).
Comparison of the predicted excess surface tension at the air/water interface, $\Delta\gamma$, from eq.~(\ref{e15})
with experimental data from Ref.~\cite{NaExp}, as function of ionic concentration, $n_b$.
The values of the $\alpha$ fitting parameter for the various salts are:   $0.179\,\kbt$ for NaF,
$0.135\,\kbt$ for NaCl,   $0.069\,\kbt$ for NaBr, and $0.023\,\kbt$ for NaI.
Other parameters are $T=300$\,K, $\varepsilon_{\rm w}=80$ (water) and $\varepsilon_{\rm a}=1$ (air).
The bottom dashed line represents the OS surface tension~\cite{onsager_samaras}. The $a$ values used in the fits
are taken from the literature and given in the text.
}}\label{fig1}
\end{figure}

\section{Surface Tension}
In order to  calculate the surface tension, we need to obtain the Helmholtz free-energy $F=\Omega+\sum_{i}\mu_i N_i$. A useful simplification for symmetric electrolytes is to replace the fugacities, $\lambda_\pm$ by the bulk densities, $n_\pm^{(b)}=n_b$. This simplification is exact on the one-loop order for $F$ (but not for $\Omega$)~\cite{tobepub}. In order to separate the volume and surface contributions in the free energy, we take explicitly the $\Lambda\rightarrow\infty$ in the first term of eq.~(\ref{e12}), obtaining the well-known Debye-H\"uckel volume fluctuation term~\cite{Debye1923}.  The final expression for the Helmholtz free-energy is then cast as
\begin{eqnarray}
\label{e13}
\nonumber F \!\!\!\!&=&\!\!\!\! \Omega_0 + 2\kbt Vn_b\ln (n_b a^3) - \frac{V\kbt}{12\pi}\kd^3 \\
&+&\!\!\!\! \frac{A\kbt}{4\pi} \left(\int_0^{\Lambda}\D k k\ln\left[\frac{\varepsilon_{\rm w} p + \varepsilon_{\rm a} k + \omega}
{\left(\varepsilon_{\rm w} + \varepsilon_{\rm a} \right)k} \right] -\frac{\omega\Lambda}{\varepsilon_{\rm w}+\varepsilon_{\rm a}}\right). \qquad
\end{eqnarray}
We note that in the above equation we already explicitly subtracted the ions volume self-energy, while the last term is the self-energy of the ions only on the surface. Expanding the last two terms in powers of $\Lambda$ yields a leading asymptotic behavior that will be discussed below.

We proceed by calculating $\Delta\gamma $, the added contribution of the electrolyte to the
surface tension of pure water at the air/water interface.
\begin{equation}
\label{e14}
\Delta\gamma \equiv \left[F - F^{(B)}(L) - F^{\rm (air)}(L)\right]/A\,,
\end{equation}
with $F^{(B)}$ being the free energy of a slab of length $L$ containing aqueous solution,
and $F^{\rm (air)}$ is the free-energy of a slab of air
\footnote{$F^{(B)}$ is obtained from eq.~(\ref{e13}) by replacing the air phase at $z<0$ with an aqueous solution,
$\varepsilon_{\rm a}\to \varepsilon_{\rm w}$. Because there is no interface at $z=0$, the ion-surface interaction vanishes leading to $\omega=0$. The only changes needed in order to obtain $F^{(B)}$ are to replace $D_1(k)/D_0(k)$ by $p/k$ and insert $\Omega_0=-2n_b$, where for the latter we use the MF solution of $\psi=0$ for bulk solvent.
Because there are no ions in the air phase, its free energy vanishes, $F^{\rm (air)}=0$.}

\begin{figure*}[ht]
\center
\includegraphics[scale=0.6,draft=false]{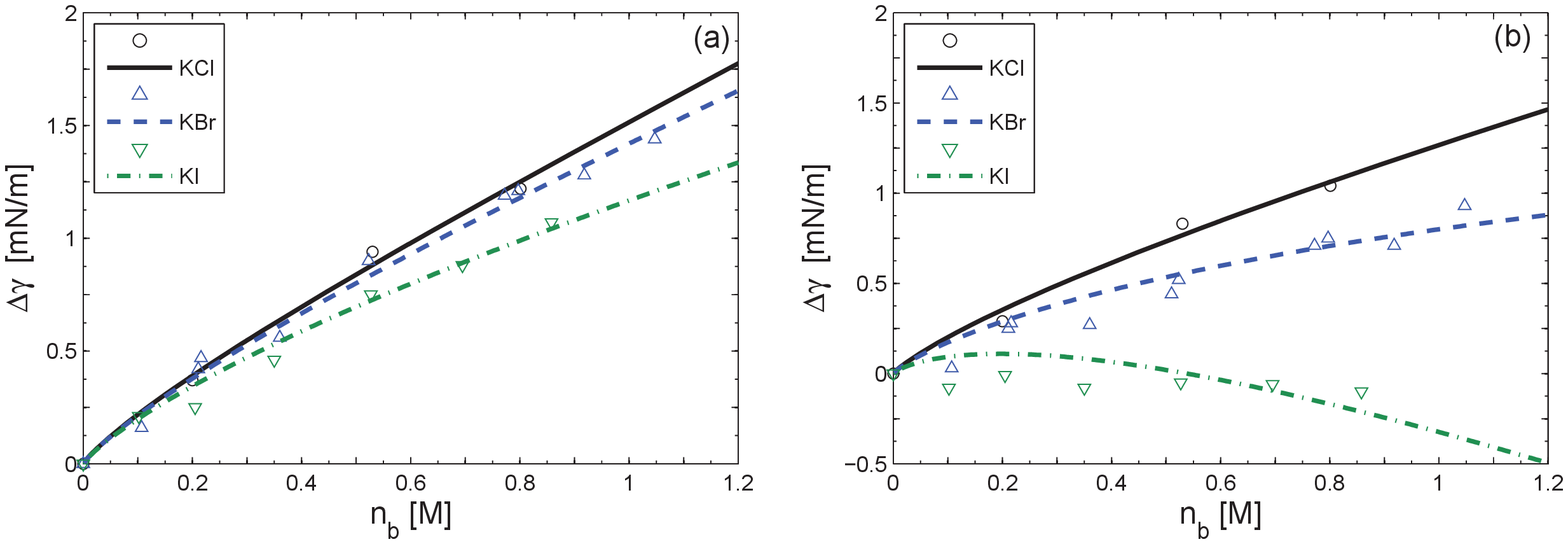} 
\caption{\textsf{(color online).
Comparison of the predicted excess surface tension, $\Delta\gamma$, from eq.~(\ref{e15})
with experimental data from Ref.~\cite{KExp}, as function of ionic concentration, $n_b$,
for the air/water interface (a) and dodecane/water (b). In (a) the fitting values of the $\alpha$ parameter for
the various salts are:  $0.137\,\kbt$ for KCl, $0.115\,\kbt$ for KBr, and $0.057\,\kbt$ for KI.
In (b) $0.085\,\kbt$ for KCl, $-0.025\,\kbt$ for KBr, and $-0.291\,\kbt$ for KI. All other parameters are as in Fig.~1, beside the dielectric constant of dodecane, $\varepsilon_d=2$.}}
\label{fig2}
\end{figure*}

The excess surface tension to one-loop order can now be written as a sum of two terms,
$\Delta\gamma=\Delta\gamma_0+\Delta\gamma_1$, and constitutes our primary analytical result:
\begin{eqnarray}
\label{e15}
\nonumber \Delta\gamma_0 &=& \kbt \left[ \frac{\sigma_{0}}{e}\,\e^{\beta e\psi_0}
- n_b \kappa_D^{-1} \left(\frac{e\psi_0}{\kbt}\right)^2\right] \\
\nonumber \Delta\gamma_1 &=& \frac{\kbt}{8\pi} \int_0^{\Lambda}\D k\, k\ln\left[\frac{k}{p}
\left(\frac{\varepsilon_{\rm w} p + \varepsilon_{\rm a} k + \omega}
{\left(\varepsilon_{\rm w} + \varepsilon_{\rm a} \right)k}\right)^2 \right] \\
& & - \, \frac{\kbt}{4\pi}\frac{\omega\Lambda}{\varepsilon_{\rm w}+\varepsilon_{\rm a}} \, .
\end{eqnarray}
The first term $\Delta\gamma_0$ is the MF excess~\cite{diamant1996}
and $\Delta\gamma_1$ is the fluctuation term, which contains the OS result~\cite{onsager_samaras,podgornik1988,dean2004} and a correction. The above result leads to an interesting observation due to the dominance of fluctuations.
As long as $\sigma_{0}$ is small, the MF term, $\Delta\gamma_{0}$, is small and the dominant
contribution comes from the fluctuation term, $\Delta\gamma_{1}$.
This observation goes hand in hand with the fact that the OS result by itself originates from fluctuations beyond MF.

The fits to the experimental data are done by evaluating the integral in eq.~(\ref{e15}) numerically for any value of $\Lambda$,
but the integral has a leading asymptotic behavior that can be obtained analytically in the $\Lambda\to\infty$ limit. Writing down only the remaining $\Lambda$-dependent terms, we obtain
\begin{eqnarray}
\nonumber \frac{8\pi}{\kbt}\Delta\gamma_{1}\!\!\!\! &\simeq&\!\!\!\! - \left(\frac{\varepsilon_{\rm w}-\varepsilon_{\rm a}}
{\varepsilon_{\rm w}+\varepsilon_{\rm a}}\right)\frac{\kappa_D^2}{2} \left[ \ln\left(\frac{1}{2}\kd\lb\right)
- \ln\left(\frac{1}{2}\lb\Lambda \right)\right. \nonumber\\
& &-  \left. \frac{2\omega^2}{\kappa_D^2(\varepsilon_{\rm w}^2-\varepsilon_{\rm a}^2)} \ln\left(\kappa_D\Lambda^{-1}\right) \right]\,.
\end{eqnarray}
The first term in $\Delta\gamma_{1}$  is the well-known OS result~\cite{onsager_samaras,podgornik1988,dean2004} and it varies as $\sim\kappa_D^2\ln(\kappa_D\lb)$,
the second term is a correction due to the ion minimal distance of approach with $\Lambda=2\sqrt{\pi}/a$, while the third term
is a correction related to the ion-surface interaction, $\alpha$.
In the limit $|\alpha| \to 0$, the latter term vanishes and the derived surface tension agrees well with
the OS result, as expected.

\section{Comparison with Experiments}

We now compare our result for the surface tension, eq.~(\ref{e15}), with experimental values~\cite{NaExp}
for four different ionic solutions (with Na$^+$ as their cation) at the air/water interface as shown in Fig.~1.
Taking $a$ as the average minimal distance between cations and anions, $a = r_{+}^{\rm hyd} + r_{-}^{\rm hyd}$,
and treating $\alpha$ as a fit parameter, we obtain very good fits to experimental data.
The values we used for $a$ are obtained from the hydrated ionic radii in Ref.~\cite{IonRadii}:
$a_{_{\rm NaF}}{=}7.1$\,\AA; $a_{_{\rm NaCl}}{=}6.9$\,\AA; $a_{_{\rm NaBr}}{=}6.88$\,\AA; $a_{_{\rm NaI}}{=}6.89$\,\AA;
$a_{_{\rm KCl}}{=}6.63$\,\AA; $a_{_{\rm KBr}}{=}6.61$\,\AA; $a_{_{\rm KI}}{=}6.62$\,\AA.
For the larger anions (with respect to their crystallographic size) Br$^{-}$ and I$^{-}$ the fit agrees well for the entire concentration range up to $\sim$1\,M,
while for the smaller anions, F$^-$ and Cl$^-$, deviations at  concentrations larger than $0.8$\,M are noticed.

Our model can be applied successfully to other types of liquid interfaces such as oil/water.
In Fig.~2, we compare a fit for air/water in (a) with oil/water in (b),
where in the experiments dodecane is used as the oil.
The fits for both interfaces are done
for the same series of three different salts having in common the K$^+$ cation,
and are in very good agreement with experiments. The only exception is the KI case at the oil/water interface,
which shows a very small $\Delta\gamma$  contribution that is almost independent of the salt concentration and, hence, is harder to fit.

We discuss now the values of $\alpha$ for seven different salts at the air/water interface as is presented in Fig. 1
(with Na$^+$ as cation) and in Fig. 2(a) (with K$^+$ as cation). They are all positive (repulsive) and range between
$ 0.02 \kbt$ and $ 0.18 \kbt$, where
$\alpha_{\rm F} > \alpha_{\rm Cl} > \alpha_{\rm Br} > \alpha_{\rm I}$ is obeyed
for both cations, and reproduces exactly the reversed Hofmeister series.
Note that the values of $\alpha$ are smaller in NaX solutions than in KX solutions (for the same X anion).
This small effect can be explained by a different ion-surface interaction of K$^+$ and Na$^+$.
We intend to further investigate this effect by introducing an extra adhesivity parameter for cations~\cite{tobepub}.
The positive values of $\alpha$ are in agreement with the values obtained by Netz and co-workers~\cite{Netz2012}, but in contrast with the effective attraction to the surface presented by Levin and co-workers\cite{levin200y}. Although in Ref.~\cite{levin200y} an effective attraction (similar to the adhesivity $\alpha$) was obtained, the trend is the same as ours, where $\alpha_{\rm F} > \alpha_{\rm Cl} > \alpha_{\rm Br} > \alpha_{\rm I}$.

At the oil/water interface [Fig.~2(b)] the same reversed Hofmeister series emerges but with a more attractive ion-surface interaction. The adhesivity $\alpha$
decreases and even becomes negative for some of the electrolytes. Since our model treats the ion-surface interaction
on a phenomenological level, we can model both attraction or repulsion of anions from the interface.
The difference in adhesivity between the air/water and oil/water interfaces is denoted by $\Delta \alpha=\alpha({\rm a/w})-\alpha({\rm o/w)}$.
The obtained $\Delta\alpha$ is different for each anion, where $\Delta\alpha_{\rm I} > \Delta\alpha_{\rm Br} > \Delta\alpha_{\rm Cl}$, and can be explained by a change in the water-surface interaction.

The present work offers several important and unique advantages. It is a self-consistent theory that extends the OS result,
and can be used quite generally for a wide variety of interfaces and surface interactions,
all taken on a common and unified ground.
The model predicts analytically the dependence of the excess surface tension of
different electrolytes at the air/water as well as at the oil/water interface.
The obtained fits agree well with experiments and show clearly the
reversed Hofmeister series (F$^{-}>$\,Cl$^{-}>$\,Br$^{-}>$\,I$^{-}$) for surface tension at both the air/water and oil/water interfaces.
It is of importance to remark that for the system parameters considered here, fluctuations
dominate over the MF contribution to the computed surface tension.

The image charge interactions are taken into
account self-consistently, hence $\alpha$ originates only from solvent structure-driven interactions
and there is no double counting.
As was discussed recently in Ref.~\cite{Sivan2009,Sivan2013} for the special case of silica/water interface,
the orientation of water molecules in the vicinity of the interface
may change the hydrogen bond strength at the interface. This surface effect can be identified as a possible microscopic
source of $\alpha$, whose value
is proportional to the difference in solvation free energy between a single ion  in the bulk  and at the surface.

Finally, it will be of interest to generalize our model to calculate surface tension
at the interface between two immiscible electrolyte solutions~\cite{Onuki_curr_op,Urbakh2006},
where the ions are present in both solutions, as well as between a variety of hydrophobic and hydrophilic solid substrates
in contact with an electrolyte solution~\cite{Sivan2009,Sivan2013}.

\acknowledgments
{\bf Acknowledgements.~~~} We thank D. Ben-Yaakov,  H. Diamant, R. Netz, H. Orland and Y. Tsori
for useful discussions and numerous suggestions.
One of us (RP) would like to thank Tel Aviv University for its hospitality during his stay there, and acknowledges the support of the ARRS through grant P1-0055.
This work was supported in part by the Israel Science Foundation (ISF) under Grant No. 438/12 and the US-Israel Binational Science Foundation (BSF) under Grant No. 2012/060.




\begin{thebibliography}{0}

\bibitem{Adamson}
\Name{A. W. Adamson \and A. P. Gast}
\Book{Physical Chemistry of Surfaces, 6th ed.}
\Publ{Wiley, New York}
\Year{1997}.


\bibitem{Pugh}
\Name{P. K. Weissenborn \and R. J. Pugh}
\REVIEW{J. Coll. Interface Sci.}{186}{1996}{550}.


\bibitem{Wagner}
\Name{C. Wagner}
\REVIEW{Phys. Z.}{25}{1924}{474}.


\bibitem {Debye1923}
\Name{P.~W. Debye \and E. H\"uckel}
\REVIEW{Phys. Z.}{24}{1923}{185}.


\bibitem{onsager_samaras}
\Name{L. Onsager \and N. N. T. Samaras}
\REVIEW{J. Chem. Phys.}{2}{1934}{628}.

\bibitem{Kunz_Book}
\Name{W. Kunz}
\Book{Specific Ion Effects}
\Publ{World Scientific, Singapore}
\Year{2009}.


\bibitem{Dan2011}
\Name{D. Ben-Yaakov, D. Andelman, R. Podgornik \and D. Harries}
\REVIEW{Curr. Opin. Coll. \& Interface Sci.}{16}{2011}{542}.

\bibitem{hofmeister}
\Name{W. Kunz, J. Henle \and B.W. Ninham}
\REVIEW{Curr. Opin. Coll. \& Interface Sci.}{9}{2004}{19}.

\bibitem{collins1985}
\Name{K. D. Collins \and M. W. Washabaugh}
\REVIEW{Q. Rev. Biophys.}{18}{1985}{323}.

\bibitem{ruckenstein2003a}
\Name{M. Manciu \and E. Ruckenstein}
\REVIEW{Adv. Colloid Interface Sci.}{105}{2003}{63}.

\bibitem{kunz2010}
\Name{W. Kunz}
\REVIEW{Curr. Opin. Coll. Interface Sci.}{15}{2010}{34}.

\bibitem{Sivan2009}
\Name{M. Dishon, O. Zohar \and U. Sivan}
\REVIEW{Langmuir}{25}{2009}{2831}.

\bibitem{Sivan2013}
\Name{J. Morag, M. Dishon \and U. Sivan}
\REVIEW{Langmuir}{29}{2013}{6317}.


\bibitem{pashely}
\Name{R. M. Pashley}
\REVIEW{J. Coll. Interface Sci.}{83}{1981}{531}.

\bibitem{parsegian1992}
\Name{D. C. Rau \and V. A. Parsegian}
\REVIEW{Biophys. J.}{61}{1992}{260}.

\bibitem{parsegian1994}
\Name{R. Podgornik, D. Rau \and V. A. Parsegian}
\REVIEW{Biophys. J.}{66}{1994}{962}.

\bibitem{air_water_2}
\Name{F. A. Long \and G. C. Nutting}
\REVIEW{J. Am. Chem. Soc.}{64}{1942}{2476}.

\bibitem{air_water_3}
\Name{J. Ralston \and T. W. Healy}
\REVIEW{J. Coll. Interface Sci.}{42}{1973}{1473}.

\bibitem {Frumkin1924}
\Name{A. Frumkin}
\REVIEW{Z. Physik. Chem.}{109}{1924}{34}.

\bibitem {dean2004}
\Name{D. S. Dean \and R. R. Horgan}
\REVIEW{Phys. Rev. E}{69}{2004}{061603}.

\bibitem {dean2003}
\Name{D. S. Dean \and R. R. Horgan}
\REVIEW{Phys. Rev. E}{68}{2003}{051104}.

\bibitem{levin200x}
\Name{Y. Levin}
\REVIEW{Phys. Rev. Lett.}{102}{2009}{147803}.

\bibitem{levin200y}
\Name{Y. Levin, A. P. dos Santos \and A. Diehl}
\REVIEW{Phys. Rev. Lett.}{103}{2009}{257802}.

\bibitem{Netz2010}
\Name{N. Schwierz, D. Horinek \and  R. R. Netz}
\REVIEW{Langmuir}{26}{2010}{7370}.

\bibitem{Netz2012}
\Name{N. Schwierz \and  R. R. Netz}
\REVIEW{Langmuir}{38}{2012}{3881}.

\bibitem{Netz2013}
\Name{N. Schwierz, D. Horinek \and  R. R. Netz}
\REVIEW{Langmuir}{29}{2013}{2602}.

\bibitem{podgornik1988}
\Name{R. Podgornik \and B. Zeks}
\REVIEW{J. Chem. Soc.}{84}{1988}{611}.

\bibitem{US1}
\Name{D. Harries, R. Podgornik, V. A. Parsegian, E. Mar-Or \and D. Andelman}
\REVIEW{J. Chem. Phys.}{124}{2006}{224702}.

\bibitem{US2}
\Name{D. Ben-Yaakov, D. Andelman, D. Harries \and R. Podgornik}
\REVIEW{J. Phys. Chem. B}{113}{2009}{6001}.

\bibitem{Onuki_curr_op}
\Name{A. Onuki \and R. Okamoto}
\REVIEW{Curr. Opin. Coll. \& Interface Sci.}{16}{2011}{525}.

\bibitem{delacruz2012}
\Name{V. Jadhao, F. J. Solis \and M. Olvera de la Cruz}
\REVIEW{Phys. Rev. Lett.}{109}{2012}{223905}.

\bibitem{diamant1996}
\Name{H. Diamant \and D. Andelman}
\REVIEW{J. Phys. Chem.}{100}{1996}{13732}.

\bibitem{tobepub}
\Name{T. Markovich, D. Andelman \and R. Podgornik}
\REVIEW{in preperation}{}{}{}.

\bibitem{Davies}
\Name{J. T. Davies}
\REVIEW{Proc. Roy. Soc. A}{245}{1958}{417}.

\bibitem{podgornik1989}
\Name{R. Podgornik }
\REVIEW{J. Chem. Phys.}{91}{1989}{9}.

\bibitem{attard}
\Name{P. Attard, D. J. Mitchell \and B. W. Ninham}
\REVIEW{J. Chem. Phys.}{88}{1987}{4987}.

\bibitem{fdet}
\Name{K. Kirsten and A. J. McKanem}
\REVIEW{Annals of Physics}{308}{2003}{502}.


\bibitem{NaExp}
\Name{N. Matubayasi, K. Tsunemoto, I. Sato, R. Akizuki, T. Morishita, A. Matuzawa \and Y. Natsukari}
\REVIEW{J. Coll. Interface Sci.}{243}{2001}{444}.

\bibitem{KExp}
\Name{R. Aveyard \and S. M. Saleem}
\REVIEW{J. Chem. Soc.}{72}{1976}{1609}.

\bibitem{IonRadii}
\Name{E. R. Nightingale Jr.}
\REVIEW{J. Phys. Chem.}{63}{1959}{1381}.
\bibitem{Urbakh2006}

\Name{C. W. Monroe, L. I. Daikhin, M. Urbakh \and A. A. Kornyshev}
\REVIEW{Phys. Rev. Lett.} {97}{2006}{136102}.


\end{thebibliography}
\end{document}